\documentstyle[amssymb,aps,12pt]{revtex}

\begin{document}
\draft
\author{O. B. Zaslavskii}
\address{Department of Mechanics and Mathematics, Kharkov V.N. Karazin's National\\
University, Svoboda\\
Sq.4, Kharkov 61077, Ukraine\\
E-mail: aptm@kharkov.ua}
\title{Exactly solvable models in 2D semiclassical dilaton gravity and extremal
black holes}
\maketitle

\begin{abstract}
Previously known exactly solvable models of 2D semiclassical dilaton gravity
admit, in the general case, only non-extreme black holes. It is shown that
there exist exceptional degenerate cases, that can be obtained by some
limiting transitions from the general exact solution, which include, in
particular, extremal and ultraextremal black holes. We also analyze
properties of extreme black holes without demanding exact solvability and
show that for such solutions quantum backreaction forbids the existence of
ultraextreme black holes. The conditions, under which divergencies of
quantum stresses in a free falling frame can disappear, are found. We derive
the closed equation with respect to the metric as a function of the dilaton
field that enables one, choosing the form of the metric, to restore
corresponding Lagrangian. It is demonstrated that exactly solvable models,
found earlier, can be extended to include an electric charge only in two
cases: either the dilaton-gravitation coupling is proportional to the
potential term, or the latter vanishes. The second case leads to the
effective potential with a negative amplitude and we analyze, how this fact
affects the structure of spacetime. We also discuss the role of quantum
backreaction in the relationship between extremal horizons and the branch of
solutions with a constant dilaton.
\end{abstract}



\section{Introduction}

Two-dimensional (2D) theories of dilaton gravity is a remarkable tool for
better understanding the role of quantum effects in black hole physics in a
more realistic four-dimensional case. The papers \cite{callan} gave a
powerful incentive to detailed studying solutions within such theories and
their properties (for recent review see, for example, \cite{od}, \cite{dv}).
As long as physical processes are being investigated perturbatively, with a
fixed classical metric chosen as the main approximation, solutions of field
equations can be found directly. As regards self-consistent solutions, they
can be always found in the classical limit since, as is known, any classical
2D gravitation-dilaton is exactly integrable. (The issue of exact solutions
becomes, however, non-trivial even classically if there is a scalar field
coupled to matter \cite{fil1}, \cite{fil2} (see \cite{eliz2} for the issue
of normalizablity) or higher order terms in curvature are taken into account 
\cite{eliz4}. Yang-Mills and fermion fields can be also included into the
scheme to give exact integrability \cite{strobl}, \cite{pelzer}. However, we
will not consider such cases.) Meanwhile, physical demands (for instance,
the necessity to trace the process of black hole evaporation with account
for quantum backreaction that may become significant at its late stage) make
it important to find exact solutions in the closed form for a {\it %
self-consistent} {\it semiclassical} (one-loop) problem. It turned out,
however, that even in spite of relative simplicity of 2D theories, such
exact solutions of semiclassical theory, in contrast to the classical one,
may be found only for some special classes of Lagrangians \cite{rst} - \cite
{exact}.

The typical features of exact solutions found in these papers consist, in
particular, in that corresponding black holes are nonextreme only and,
moreover, the Hawking temperature is constant in the sense it does not
depend on a horizon value of the dilaton. Apart from this, all exactly
solvable models discussed in \cite{rst} - \cite{exact} refer to the
uncharged case. However, the issue of extreme black holes, in connection
with its obvious importance (for example, as potential candidates on the
role of stable remnants after black hole evaporation), makes it actual to
elucidate, whether it is possible to extend the approach to solvability from
the uncharged case to the charged one to embrace the extreme case. Besides,
even for uncharged black holes, it may happen that the solutions considered
in \cite{rst} - \cite{exact} do not exhaust all possibilities and some other
exact solutions for extreme black holes are also possible. We address these
issues below. Meanwhile, the question of expanding classes of exactly
solvable models in 2D dilaton gravity deserves treatment by itself, not only
in the context of extreme black hole solutions. Thus, our aim is triple: (i)
to expand families of exactly solvable models, (ii) take into account an
electric charge, (iii) elucidate the conditions, under which extreme black
holes exist in 2D semiclassical theories (in generic and exactly solvable
models), and compare this situation with that in the classical theory.

The paper is organized as follows. In Sec. II we list field equations and
the condition of exact solvability according to \cite{exact}. In Sec. III we
suggest alternative derivation of this condition, especially natural in the
context of black hole solutions. In Sec. IV we consider the special class of
solution with a constant dilaton field which does not fall into a general
scheme. Sec. V is devoted to classical charged black holes where we
generalize some known examples. In Sec. VI it is shown that some non-trivial
limiting transitions in models of \cite{exact} give rise to qualitatively
new possibilities, including the existence of extreme black holes. In Sec.
VII we relax the condition of exact solvability, analyze generic extreme
black holes and examine the possibility of ultraextreme quantum-corrected
black holes. Sec. VIII summarizes the main results.

\section{Basic equations}

Let us consider the system governed by the action 
\begin{equation}
I=I_{0}+I_{PL}\text{, }I_{0}=I_{gd}+I_{q}\text{,}  \label{1}
\end{equation}
where gravitation-dilaton part 
\begin{equation}
I_{gd}=\frac{1}{2\pi }\int_{M}d^{2}x\sqrt{-g}[F(\phi )R+V(\phi )(\nabla \phi
)^{2}+U(\phi )]\text{,}  \label{2}
\end{equation}
the contribution of the electromagnetic field is 
\begin{equation}
I_{q}=-\frac{1}{4\pi }\int_{M}d^{2}x\sqrt{-g}W(\phi )F_{\mu \nu }F^{\mu \nu }%
\text{,}
\end{equation}
$F_{\mu \nu }F^{\mu \nu }=2F_{01}F^{01}\equiv -2E^{2}$ and we omit boundary
terms that does not affect the form of field equations.

$I_{PL\text{ }}$is the Polyakov-Liouville action \cite{polyakov}
incorporating effects of Hawking radiation and its backreaction on spacetime
for a multiplet of N conformal scalar fields. It is convenient to write it
down in the form \cite{solod}, \cite{israel} 
\begin{equation}
I_{PL}=-\frac{\kappa }{2\pi }\int_{M}d^{2}x\sqrt{-g}[\frac{(\nabla \psi )^{2}%
}{2}+\psi R]-\frac{\kappa }{\pi }\int_{\partial M}\psi kds\text{.}  \label{3}
\end{equation}
The function $\psi $ obeys the equation 
\begin{equation}
\square \psi =R\text{,}  \label{4}
\end{equation}
where $\Box =\nabla _{\mu }\nabla ^{\mu }$, $\kappa =N/24$ is the quantum
coupling parameter. Varying the action with respect to the metric, we get 
\begin{equation}
T_{\mu \nu }\equiv T_{\mu \nu }^{(gd)}+T_{\mu \nu }^{(q)}+T_{\mu \nu
}^{(PL)}=0\text{,}  \label{6}
\end{equation}
where 
\begin{equation}
T_{\mu \nu }^{(gd)}=\frac{1}{2\pi }\{2(g_{\mu \nu }\square F-\nabla _{\mu
}\nabla _{\nu }F)-Ug_{\mu \nu }+2V\nabla _{\mu }\phi \nabla _{\nu }\phi
-g_{\mu \nu }V(\nabla \phi )^{2}\}\text{,}  \label{7}
\end{equation}
\begin{equation}
T_{\mu \nu }^{(PL)}=-\frac{\kappa }{2\pi }\{\partial _{\mu }\psi \partial
_{\nu }\psi -2\nabla _{\mu }\nabla _{\nu }\psi +g_{\mu \nu }[2\square \psi -%
\frac{1}{2}(\nabla \psi )^{2}]\}\text{,}  \label{8}
\end{equation}
\begin{equation}
T_{\mu \nu }^{(q)}=-\frac{W}{2\pi }(g_{\mu \nu }E^{2}+2g^{\alpha \beta
}F_{\mu \alpha }F_{\nu \beta })\text{.}  \label{tq}
\end{equation}
Variation of the action with respect to $\phi $ gives rise to the equation 
\begin{equation}
RF^{^{\prime }}+U^{\prime }=2V\Box \phi +V^{\prime }(\nabla \phi )^{2}\text{,%
}  \label{9}
\end{equation}
where prime denotes derivative with respect to $\phi $.

The Maxwell equations read 
\begin{equation}
(WF^{\mu \nu })_{;\nu }=0\text{.}
\end{equation}
The electromagnetic field in our 2D case is antisymmetric: 
\begin{equation}
F_{\mu \nu }=-Ee_{\mu \nu }\text{,}
\end{equation}
where $e_{\mu \nu }=-e_{\mu \nu }$ and $e_{01}=(-g)^{1/2}$. Then it follows
from (\ref{tq}) that 
\begin{equation}
T_{\mu }^{\nu (q)}=\frac{WE^{2}}{2\pi }\delta _{\mu }^{\nu }\text{.}
\end{equation}

Let us suppose that the auxiliary field $\psi $ depends on one variable $%
\phi $ only. Then the structure of field equations exhibits the existence of
the Killing vector $\xi _{\alpha }=e_{\alpha }^{\beta }\mu _{;\beta }$,
where $\mu ^{\prime }=\tilde{F}^{\prime }\exp (-\int d\phi \frac{\tilde{V}}{%
\tilde{F}^{\prime }})$,

\begin{equation}
\tilde{F}=F-\kappa \psi \text{, }\tilde{V}=V-\kappa \frac{\psi ^{\prime 2}}{2%
}\text{.}  \label{shift}
\end{equation}
It can be either spacelike or timelike. In what follows we restrict
ourselves to the static case (timelike Killing vector). As in the static
case both $\psi $ and $\phi $ depend on the spatial coordinate $x$ only ($%
\psi =\psi (x)$ and $\phi =\phi (x)$), we may exclude $x$ and express $\psi $
in terms of $\phi $ directly, so our assumption $\psi =\psi (\phi )$ is
self-consistent and is valid for any static space-time.

Thus, we consider space-times which have no dynamics. Although, formally,
radiative solutions do not fall into this class, the models under
consideration are intimately related to them. To describe processes of
formation and evaporation of black holes, we should add to
gravitation-dilaton part of the action also classical matter fields, which
in the simplest case represent shock waves. Then the systems we are dealing
with in the present paper may be used to describe different parts of
space-time before and after shock wave - an intial state, remnants that
represent the result of evaporation, etc. (see, for instance, careful
analysis of dynamic scenario of one particular exactly solvable model in 
\cite{bose}). Anyway, the description of static solutions and their
properties (to which we restrict ourselves in the present article) is an
essential ingredient for the full analysis of a more complex dynamic picture.

Let us return to the issue of field equations. Now the Gauss law takes the
form ($\mu =0$) 
\begin{equation}
(W\sqrt{-g}F^{01})_{,1}=0\text{,}
\end{equation}
whence $F_{01}=-QW^{-1}\sqrt{-g}$, $E^{2}=Q^{2}W^{-2}$, where the constant $%
Q $ has physical meaning of an electric charge. It follows from (\ref{tq})
that the dilaton equation reads 
\begin{equation}
RF^{^{\prime }}+U^{\prime }=2V\square \phi +V^{\prime }(\nabla \phi
)^{2}-E^{2}W^{\prime }\text{,}  \label{dil}
\end{equation}

From the trace equations we obtain 
\begin{equation}
U=\square \tilde{F}+WE^{2}\text{,}  \label{u}
\end{equation}
\begin{equation}
A_{1}\square \phi +A_{2}(\nabla \phi )^{2}+E^{2}(\omega W+W^{\prime })=0%
\text{,}  \label{bas}
\end{equation}
\begin{equation}
A_{1}=(u-\kappa \omega )\psi ^{\prime }+\omega u-2V\text{,}  \nonumber
\end{equation}
\[
A_{2}=(u-\kappa \omega )\psi ^{\prime \prime }+\omega u^{\prime }-V^{\prime }%
\text{,} 
\]
where $u\equiv F^{\prime }$, $\omega \equiv \frac{U^{\prime }}{U}$.

It is worth noting that in the presence of an electric charge field
equations retain the form, typical of the uncharged case, but with the
redefinition $U\rightarrow U_{eff}$, where 
\begin{equation}
U_{eff}\equiv U-Q^{2}W^{-1}\text{.}  \label{ueff}
\end{equation}

If an electromagnetic field is absent ($E=0)$, the condition of exact
solvability looks like \cite{exact} 
\begin{equation}
V=\omega (u-\frac{\kappa \omega }{2})+C(u-\kappa \omega )^{2}\text{.}
\label{vex}
\end{equation}

This condition was derived (with the absence of an electromagnetic field) in 
\cite{exact} from the demand that eq. (\ref{bas}) not contain gradient of
the dilaton field, so that $A_{1}=A_{2}=0$. Meanwhile, before considering
the situation with an electromagnetic field, it is instructive to look at
the exact solvability in the case $E=0$ from another viewpoint.

\section{New approach to exact solvability}

Let us write down the metric in the Schwarzschild gauge: 
\begin{equation}
ds^{2}=-fdt^{2}+f^{-1}dx^{2}  \label{gauge}
\end{equation}
It is worth noting that field equations of the semiclassical system under
discussion look very much like those for a pure classical, but with
coefficients, shifted according to (\ref{shift}).

Then field equations take the following explicit form: 
\begin{equation}
2f\frac{\partial ^{2}\tilde{F}}{\partial x^{2}}+\frac{\partial f}{\partial x}%
\frac{\partial \tilde{F}}{\partial x}-U-\tilde{V}f\left( \frac{\partial \phi 
}{\partial x}\right) ^{2}=0\text{,}  \label{sc00}
\end{equation}

\begin{equation}
\frac{\partial f}{\partial x}\frac{\partial \tilde{F}}{\partial x}-U+\tilde{V%
}f\left( \frac{\partial \phi }{\partial x}\right) ^{2}=0\text{.}
\label{sc11}
\end{equation}
In our gauge (\ref{gauge}) the integration of eq. (\ref{4}) gives us 
\begin{equation}
\frac{\partial \psi }{\partial x}=\frac{A-f_{x}^{\prime }}{f}.  \label{ps}
\end{equation}

Let $\frac{\partial \phi }{\partial x}\equiv z(\phi )$. It is also
convenient to take the difference of (\ref{sc00}), (\ref{sc11}) to get 
\begin{equation}
\frac{\partial ^{2}\tilde{F}}{\partial x^{2}}=\tilde{V}\left( \frac{\partial
\phi }{\partial x}\right) ^{2}\text{.}  \label{dif}
\end{equation}
Then it follows that 
\begin{equation}
f=\chi \exp (-\rho )\text{, }\chi ^{\prime }=\frac{U}{z^{2}\tilde{u}}e^{\rho
}\text{, }\rho =\int d\phi \alpha \text{, }\alpha \equiv \frac{\tilde{V}}{%
\tilde{u}}\text{, }\tilde{u}=u-\kappa \psi ^{\prime }\text{.}  \label{ff}
\end{equation}
\begin{equation}
\tilde{V}=z^{-1}\frac{\partial }{\partial \phi }(\tilde{u}z)=\tilde{u}%
^{\prime }+\tilde{u}\frac{z^{\prime }}{z}\text{.}  \label{v}
\end{equation}
Inverting (\ref{v}), we get 
\begin{equation}
z=z_{0}\tilde{u}^{-1}e^{\rho }\text{, }z_{0}=const\text{,}  \label{qh}
\end{equation}
\begin{equation}
\chi ^{\prime }=\frac{U\tilde{u}}{z_{0}^{2}}\exp (-\rho )\text{,}  \label{hi}
\end{equation}
\[
f=\exp (-\rho )\int_{\phi _{h}}^{\phi }d\phi \chi ^{\prime }\text{,} 
\]
the point $\phi _{h}$ corresponds to the event horizon of a black hole,
whence 
\begin{equation}
x=\int d\phi z^{-1}=z_{0}^{-1}\int d\phi \tilde{u}\exp (-\rho )\text{.}
\label{coor}
\end{equation}

Multiplying (\ref{ps}) by $\frac{\partial x}{\partial \phi }$and taking into
account (\ref{ff}) - (\ref{coor}), we get 
\begin{equation}
\frac{\psi ^{\prime }-\rho ^{\prime }}{\tilde{u}}\int_{\phi _{h}}^{\phi
}d\phi ^{\prime }w(\phi ^{\prime })=\frac{w(\phi _{h})}{\tilde{u}(\phi _{h})}%
-\frac{w(\phi )}{\tilde{u}(\phi )}\text{, }w\equiv \chi ^{\prime }\text{.}
\label{int}
\end{equation}
Here we assumed that $\psi $ is regular on the horizon, whence in eq. (\ref
{ps}) the constant $A=\left( \frac{\partial f}{\partial x}\right)
_{x=x_{h}}=z_{0}\frac{w(\phi _{h})}{\tilde{u}(\phi _{h})}$.

By their very meaning, the action coefficients $F$, $U$, $V$ should not
depend on $\phi _{h}$, which is the characteristics of a particular solution
only. The same is true for the quantity $\psi (\phi )$ and, therefore, for $%
\tilde{u}(\phi )$ and $w(\phi )$ as well. Differentiating (\ref{int}) with
respect to $\phi _{h}$, we get 
\begin{equation}
\frac{(\psi ^{\prime }-\rho ^{\prime })}{\tilde{u}}w(\phi _{h})=-\frac{%
\partial }{\partial \phi _{h}}\frac{w(\phi _{h})}{\tilde{u}(\phi _{h})}\text{%
.}
\end{equation}
There are two independent variables $\phi $ and $\phi _{h}$ in this
equation. Solving it by separation of variables, we find that 
\begin{equation}
\psi ^{\prime }=\rho ^{\prime }+b\tilde{u}\text{, }
\end{equation}
\begin{equation}
\frac{w}{\tilde{u}}=U_{0}e^{-b\tilde{F}}\text{,}  \label{wu}
\end{equation}
$b$ and $U_{0}$ are constants. Comparing (\ref{wu}) with (\ref{hi}) and
remembering the definitions of $\rho $ and $\omega \equiv U^{\prime }/U$, we
find 
\begin{equation}
\tilde{V}=\omega \tilde{u}+b\tilde{u}^{2}  \label{v1}
\end{equation}
\begin{equation}
\psi ^{\prime }=\omega +2b\tilde{u}\text{.}  \label{pst}
\end{equation}
Eq. (\ref{pst}) is equivalent to (\ref{vex}), if one identifies constants $b=%
\frac{C}{1-2\kappa C}\equiv \tilde{C}$. Throughout the present paper we put
for simplicity $C=0$, so for exactly solvable models 
\begin{equation}
V=\omega u-\frac{\kappa \omega ^{2}}{2}\text{.}  \label{v0}
\end{equation}
Thus, we arrive at the conditions of exact solvability, derived in \cite
{exact} in a quite different way.

\section{Solutions with constant $\phi $}

In the preceding section it was tacitly assumed that the dilaton field is
not constant identically. Otherwise, this special case is needed to be
discussed separately. Let $\phi =\phi _{0}=const$. In terms of the
spherically-symmetrical reduction from 4D theory to 2D one, the dilaton is
analog of the coefficient $r^{2}$ at the angular part of the 4D interval.
Therefore, the solutions with a constant dilaton are analogs of $r^{2}=const$
solutions, which are nothing else than the Bertotti-Robinson (BR)
spacetimes. In doing so, the constant dilaton value corresponds to the
horizon area.

For the constant dilaton the expression for $T_{\mu \nu }$ simplify:

\begin{equation}
T_{\mu \nu }^{(gd)}=-\frac{U(\phi _{0})}{2\pi }\delta _{\mu }^{\nu }\text{, }
\end{equation}
\begin{equation}
T_{\mu \nu }^{(PL)}=-\frac{\kappa R}{\pi }\delta _{\mu }^{\nu }\text{.}
\end{equation}

We obtain from field equations, including the dilaton one: 
\begin{equation}
U_{eff}(\phi _{0})+\kappa R=0\text{,}  \label{c1}
\end{equation}
\begin{equation}
RF^{^{\prime }}(\phi _{0})+U_{eff}^{\prime }(\phi _{0})=0\text{,}  \label{c2}
\end{equation}
where $U_{eff}$ is taken from eq. (\ref{ueff}).

In particular, for the string-inspired dilaton gravity theory \cite{frolov}
with 
\begin{equation}
F=e^{\phi }=V\text{, }U=\lambda e^{\phi }\text{, }W=e^{\phi }\text{, }%
U_{eff}=\lambda e^{\phi }-Q^{2}e^{-\phi }  \label{98}
\end{equation}
our previous results \cite{const} can be obtained from (\ref{c1}), (\ref{c2}%
). It was shown that in the absence of an electromagnetic field, the
solutions with a constant dilaton in exactly solvable models are possible
only due to quantum effects \cite{solod}, \cite{exact}. However, now they
appear in the classical domain as well.

There are two qualitatively different possibilities.

1) $R=0$ (flat spacetime). This restricts the parameters of a system to some
particular values. For example, if 
\begin{equation}
U_{eff}=\lambda e^{2\phi }+\Lambda -Q^{2}e^{\phi }\text{,}  \label{u0}
\end{equation}
the solution is possible only if $Q^{4}=4\lambda \Lambda \equiv Q_{1}^{4}$,
i.e. for a singled value of charge. This means such a balance between
dilaton and electric forces, that combined action of both these sources does
not curve spacetime.

2) $R\neq 0$. Then there are three possible cases: 
\begin{equation}
ds^{2}=-dt^{2}\mu ^{2}+dl^{2}\text{,}
\end{equation}
where $\mu =$(i) $e^{-cl}$, (ii) $\frac{\sinh cl}{c}$ or (iii) $\frac{\sin cl%
}{c}$. In the cases (i), (ii) $R=-2c^{2}<0$, in (iii), $R=2c^{2}>0.$

For (\ref{98}) in the case (iii) we need, according to eqs. (\ref{c1}), (\ref
{c2}) to combine $U_{eff}^{\prime }(\phi _{0})<0$ and $U_{eff}(\phi _{0})<0$%
. This is possible only in the essentially quantum case: $\kappa >e^{\phi
_{0}}$ and for $\lambda <0$ \cite{const}. However, for (\ref{u0}) this can
be achieved even classically and with $\lambda >0$, if $Q>Q_{1}.$

In the classical case, if eq. (\ref{c1}) has more than one root, any two
successive roots correspond to the curvatures of different signs (provided $%
F^{\prime }$ does not change the sign) since derivative $U_{eff}^{\prime }$
changes its sign. With quantum backreaction taken into account the constant
dilaton values move in such a way that 
\begin{equation}
signU^{\prime }=signF^{\prime }signU\text{.}  \label{signs}
\end{equation}

It was stressed in \cite{deg} that the constant dilaton solution appears
just at the values which correspond to degenerate horizons of the basic
branch of solution ($\phi \neq const$). This conclusion was reached for
classical systems. Now we will show that account for backreaction retains
this relationship. Indeed, consider merging horizons. Then near the horizon $%
f=f_{0}(x-x_{h})^{2}+...$. To have finite quantum stresses on the horizon,
we should put $A=0$ in (\ref{ps}) that is equivalent to $T=T_{H}=0$ \cite
{and}. Then near the degenerate horizon 
\begin{equation}
\frac{\partial \psi }{\partial x}=-2(x-x_{h})^{-1}+...  \label{psh}
\end{equation}
With account for (\ref{shift}), (\ref{psh}), eqs. (\ref{sc00}) and (\ref
{sc11}) give us on the horizon the same condition $U_{eff}(\phi _{h})+\kappa
R(\phi _{h})=0$, $R(\phi _{h})=-2f_{0}$ for a basic branch (non-constant
dilaton). This condition corresponds just to (\ref{c1}) for the identically
constant dilaton. For any regular function $\phi (x)$ we have $\square \phi
=f^{\prime }\phi ^{\prime }+f\phi ^{\prime \prime }=0$ on the degenerate
horizon, whence we get from the dilaton equation (\ref{9}) the condition (%
\ref{c2}). Thus, if a basic branch admits merging horizons, the intimate
connection between both types of solutions persists in the quantum case too.

It is worth stressing, however, that this kind of correspondence between two
branches of solutions can be broken in two cases. First, when two horizons
merge they form a degenerate horizon (an extremal black hole); however, such
extremal horizons may be absent in some classes of solutions. For example,
all black holes pertaining to exactly solvable models considered in \cite
{exact} are non-extreme and constant dilaton solutions correspond to
singularities of the main branch \cite{solod}, \cite{exact}. Second, even if
the main branch (with a non-constant dilaton) does admit extremal horizon,
there exist solutions of field equations with a regular extremal horizon but
infinite quantum stresses, when $A\neq 0$ \cite{ext}, \cite{mod} (see also
Sec. VIII below). Then derivation of eq. (\ref{c1}) for corresponding
quantities on an extremal horizon of the main branch loses its validity
since the equality $A=0$ was used there in an essential way (see above). For
instance, the effective potential for extreme solutions in \cite{ext} $U=0$
on the horizon but the curvature $R\neq 0$ there, so eq. (\ref{c1}) is not
fulfilled for the main branch. Thus, quantum backreaction may give rise to
such extremal horizon which have no counterparts in the constant dilaton
solutions.

\section{Classical charged black holes}

In the classical limit, when $\kappa =0$, any 2D dilaton gravity model
becomes integrable. This fact is known. However, as some models of charged
2D dilaton black holes play an especial physical role in the context of the
string theory \cite{frolov}, \cite{gibper}, \cite{nappi} and are under
intensive study up to now \cite{gukov}, we dwell upon on this case
separately. There is an elaborated approach for obtaining solutions, which
is based on simplifying Lagrangian due to excluding the kinetic term \cite
{kunst1}. Instead, we give direct expressions for solutions in the original
conformal frame without additional manipulations.

As now tilted coefficients coincide with the original ones, eqs. (\ref{ff})-(%
\ref{coor}) give the complete solution of the problem since an unknown
function $\psi (\phi )$ does not appear in their right hand side. All
relevant formulas contain now, instead of the potential $U$, the effective
potential (\ref{ueff}).

Consider several examples.

Example 1. $V=e^{\phi }$, $F=e^{\phi }$, $U=\lambda ^{2}e^{\phi }+\Lambda $, 
$W=e^{\phi }$, $U_{eff}=$ $\lambda e^{\phi }+\Lambda -Q^{2}e^{-\phi }$. Let
us choose the constant $z_{0}=\lambda _{0}$. Then 
\begin{equation}
f=1-\frac{2M}{\lambda }e^{-\phi }+\frac{Q^{2}}{\lambda ^{2}}e^{-2\phi }+%
\frac{\Lambda \phi }{\lambda ^{2}}.  \label{fguk}
\end{equation}
In the limit $\Lambda \rightarrow 0$ the result of \cite{frolov} is
reproduced, the limit $Q\rightarrow 0$ corresponds to \cite{gukov}.

The solution for charged configurations may contain, in particular, extreme
black holes. By definitions, for such black holes the Hawking temperature $%
T_{H}=\frac{f^{\prime }(x_{W})}{4\pi }=0$ ($x_{h}$ corresponds here to the
horizon, where $f(x_{h})=0$). Let everywhere $u\neq 0$, the function $F(\phi
)$ is monotonic. Then it follows from (\ref{ff}) - (\ref{coor}) that $%
U_{eff}(\phi _{h})=0$. In principle, it may happen that in the point $\phi
_{h}$ also $U_{eff}^{\prime }(\phi _{h})=0$. Then we get an ''ultraextreme''
black holes for which near the horizon $f=const(x-x_{h})^{3}$ (by
definition, a black hole is ultraextreme if near the horizon $f\backsim
(x-x_{h})^{n}$ with $n>2$). This can be achieved if, say, we add the term of
the kind $Be^{-\phi }$ to $U$. Such solutions were impossible for \cite
{frolov}, \cite{gukov}.

Example 2. $F=e^{-2\phi }$, $V=4e^{-2\phi }$, $U=-2\lambda ^{2}\sinh 2\phi $%
\cite{cad}. Choosing the constant $z_{0}=\lambda $ to ensure $f\rightarrow 1$
at $x\rightarrow +\infty $, we get 
\begin{equation}
x=-\frac{2}{\lambda }(\phi -\phi _{0})\text{, }f=(1-e^{2\phi -2\phi
_{W}})(1-e^{2\phi +2\phi _{h}})\text{,}
\end{equation}
where $\phi _{h}$ is the horizon value. If one identifies $a_{+}=\exp (2\phi
_{0}-2\phi _{h})$, $a_{-}=\exp (2\phi _{0}+2\phi _{h})$, we get 
\begin{equation}
f=(1-a_{+}e^{-\lambda x})(1-a_{-}e^{-\lambda x})\text{,}
\end{equation}
that reproduces eq. (7) of \cite{cad}.

\section{Exact solutions for semiclassical charged black holes}

Let us now consider charged solutions in semiclassical theories of dilaton
gravity and try to generalize exactly solvable models to take into account
an electromagnetic field. Our strategy consists of two steps. First, we find
exactly solvable solutions for the effective potential $U_{eff}$ (\ref{ueff}%
). As both $U_{eff\text{ }}$ and $\omega _{eff}=U_{eff}^{\prime }/U_{eff}$
contain the charge $Q$, this charge will, in general, enter the expressions
for the functions $u(\phi )$ and $V(\phi )$. This is undesirable feature
since the charge is characteristic of a particular solutions and does not
appear in the original Lagrangian. In this sense, different exactly solvable
models would correspond, generally speaking, to different $Q$, each for one
value of $Q$ only. In other words, if some model was exactly solvable for a
fixed value of $Q$, small variation of $Q$ would destroy exact solvability.

To exclude this and find the models, exactly solvable for any $Q$, we need
to make the second step and get rid of $Q$-dependence in the action
coefficients.

It is clear from (\ref{vex}) that the only way to get rid of the $Q$%
-dependence in $V$ is to demand that $\omega _{eff}$ not contain $Q$. Using
the definition of $\omega _{eff}$, we have 
\begin{equation}
\omega _{eff}=\frac{U^{\prime }+Q^{2}W^{-2}W^{\prime }}{U-Q^{2}W^{-1}}\text{.%
}
\end{equation}
There are only two possibilities here to achieve our goal.

\subsection{Case 1)}

1) $W=W_{0}U^{-1}$, where $W_{0}=const$. Then $U_{eff}\backsim U$ and $%
\omega _{eff}=\omega =U^{\prime }/U$. In terms of eq. (\ref{bas}) this means
that the third term is identically zero. Then the analysis of \cite{exact}
(where an electromagnetic field was not taken into account) applies
directly. The only difference is that the presence of an electric charge
affects the value of the amplitude of the effective potential. If we write
the potential in the form 
\begin{equation}
U=4\lambda ^{2}\exp (\int_{0}^{\phi }d\phi \omega )\text{,}
\end{equation}
then replacement $U$ by $U_{eff}$ means the change 
\begin{equation}
\lambda \rightarrow \bar{\lambda}=\lambda \sqrt{1-\frac{Q^{2}}{W_{0}}\text{ }%
}  \label{l}
\end{equation}
(cf. \cite{park}, where, however, the system was supposed to be a pure
classical). Correspondingly, all formulas look very much like for the
uncharged case \cite{exact}, if the parameter $\lambda $ is replaced by $%
\bar{\lambda}$.

Much more interesting possibilities arise in the second case since they can
lead to qualitative changes in the structure of spacetime of exactly
solvable models which we now turn to.

\subsection{Case 2)}

Let $U=0$, $W$ $\equiv e^{\int d\phi \omega }$ be arbitrary, $%
U_{eff}=-Q^{2}W^{-1}$. Then 
\begin{equation}
\omega _{eff}=-\frac{W^{\prime }}{W}.
\end{equation}
Using the results of \cite{exact}, one can write (up to the constant factor)
the explicit expression for the solution in the form 
\begin{equation}
f=a(\tilde{F}_{h}-\tilde{F})W\text{, }\tilde{F}_{h}\equiv \tilde{F}(\phi
_{h})\text{,}  \label{fabh}
\end{equation}
\begin{equation}
\text{ }\frac{\partial x}{\partial \phi }=B^{-1}\tilde{F}^{\prime }W\text{,}
\end{equation}
where $a$ and $B$ are constants, related, according to (\ref{u}), by the
equation 
\begin{equation}
aB^{2}=Q^{2}\text{.}  \label{aB}
\end{equation}
As in the case (1), the value of charge affects not the coordinate
dependence of the metric explicitly but only the normalization factors like $%
\lambda $ and $a$.

The Riemann curvature 
\begin{equation}
R=\frac{Q^{2}}{\tilde{F}^{\prime }W}\frac{\partial }{\partial \phi }\left[ 
\frac{\left( \tilde{F}-\tilde{F}_{h}\right) W^{\prime }}{\tilde{F}^{\prime }W%
}\right] \text{.}
\end{equation}

The roots of the equation $\tilde{F}^{\prime }(\phi _{s})=0$ correspond to
timelike or spacelike singularities, provided $W(\phi _{s})\,$and $W^{\prime
}(\phi _{s})$ are finite and nonzero. If the function $W$ has a zero at $%
\phi =\phi _{1}$, this root corresponds to the singular horizon. In general,
the behavior of $W(\phi )$ near the point $\phi _{s}$ may affect the
character of singularity and even remove it, making spacetime everywhere
regular.

According to \cite{thr}, for exactly solvable semiclassical dilaton gravity
theories solutions of field equations can be written as 
\begin{equation}
\tilde{F}^{(0)}=h(\sigma )\equiv C+De^{-\sigma \delta }+\kappa \gamma (1-%
\frac{\gamma }{2\delta })\sigma \text{,}  \label{F}
\end{equation}
where $\sigma $ is a conformal coordinate, $\tilde{F}^{(0)}=F-\kappa \psi
_{0}$, 
\begin{equation}
ds^{2}=f(-dt^{2}+d\sigma ^{2})\text{,}  \label{conf}
\end{equation}
\begin{equation}
f=e^{-\psi _{0}-\delta \sigma }\text{,}  \label{f}
\end{equation}
\begin{equation}
\psi _{0}=\ln U_{eff}+const\text{,}  \label{psu}
\end{equation}
\begin{equation}
\psi =\psi _{0}+\gamma \sigma \text{,}  \label{psg}
\end{equation}
Now the potential $U$ \cite{thr} should be replaced, as is explained above,
by $U_{eff}=-Q^{2}W^{-1}$. If we write down the potential in the form $%
U\equiv \Lambda \exp (\int d\phi \omega )$, it follows from the trace
equation that for exactly solvable models (\cite{thr}) the constants obey
the equation 
\begin{equation}
D\delta ^{2}=\Lambda \text{.}  \label{ddl}
\end{equation}
For the case under discussion $\Lambda =-Q^{2}<0.$

Thus, in contrast to \cite{exact}, now $D<0$. Shifting $\sigma $, one can
achieve $D=-1$. Introducing dimensionless coordinate $y=\frac{Q\sigma }{2}$
and choosing $\delta =-Q$ (the factor $2$ is introduced to retain succession
with previous papers \cite{exact}, \cite{thr}), we get 
\begin{equation}
\tilde{F}^{(0)}=-e^{2y}-By+C\equiv h(y)\text{,}  \label{fh}
\end{equation}
$B\equiv -\kappa \frac{2\gamma }{Q}(1+\frac{\gamma }{2Q})$.

It is shown in \cite{thr} that the expression for the metric can be written
for exactly solvable models in the form

\begin{equation}
f=ae^{2y}W\text{.}  \label{fa}
\end{equation}
For black holes, the coefficient $B$ arises due to deviation of temperature $%
T$ from its Hawking value $T_{H}$ \cite{nonext}. Demand $T=T_{H}$ leads to $%
B=0$, in which case one can easily see that (\ref{fa}) is equivalent to (\ref
{fabh}).

For definiteness, consider the dilaton-electromagnetic coupling of the form $%
W=e^{2\phi }$. Then $U_{eff}=-Q^{2}e^{-2\phi }$. We choose the quantity $%
\tilde{F}$ corresponding to the CN\ \cite{cruz} model that generalizes the
RST \cite{rst} and BPP \cite{bose} ones: 
\begin{equation}
\tilde{F}^{(0)}=\exp (-2\phi )+2\kappa d\phi \text{.}  \label{eq}
\end{equation}
The amplitude of the potential is negative but, as the quantity $V$ for
exactly solvable models contain $\omega \equiv \acute{U}_{eff}^{\prime
}/U_{eff}$, the sign of the amplitude is irrelevant, and $V$ is the same as
in \cite{cruz}: 
\begin{equation}
V=4\exp (-2\phi )+2(1-2d)\kappa \text{, }U_{eff}=-Q^{2}e^{-2\phi }\text{.}
\label{vv}
\end{equation}

It is convenient to calculate the curvature in the conformal frame according
to the formula (where we have taken into account that for the model under
discussion $\omega =-2$) 
\begin{equation}
R=-\frac{Q^{2}}{2}f^{-1}\frac{\partial ^{2}\phi }{\partial y^{2}}\text{.}
\label{ry}
\end{equation}

The main difference as compared to previously considered exactly solvable
models \cite{exact}, \cite{thr}, \cite{nonext} is that now the coefficient
at $e^{2y}$ in $h(y)$ changed the sign. At $y\rightarrow -\infty $ (near the
horizon) this has a negligible effect. However, it affects the behavior of $%
h(y)$, when $y$ grows$.$ Consider several possible cases separately and
comment mainly on features of the spacetime which arise due to the negative
sign of $D$ in (\ref{F}).

1) $d=0$.

a) $B\geq 0$. Now we have a region $-\infty <y\leq y_{0}$ between a horizon
and the point $y_{0}$ such that $h(y_{0})=0$. If $h^{\prime }(y_{0})\neq 0$,
it corresponds to a singularity, if $h^{\prime }(y_{0})=0$, we have there a
semi-infinite throat.

b) $B<0$. There are two roots $h(y_{1})=h(y_{2})=0$, if $C>\frac{\left|
B\right| }{2}\ln \frac{\left| B\right| }{2e}$ (otherwise, there are no
admissible solutions at all). Near each root we have a singularity.

2) $d>0$. Now $\tilde{F}(\phi )$ has a minimum at $\phi =\phi _{0}$, $\tilde{%
F}^{(0)\prime }(\phi _{0})=0$, eq. (\ref{fh}) for each $y$ has two solutions.

a) $B\geq 0$. As $h\rightarrow -\infty $ at right infinity, $y$ cannot grow
unbounded, $y<y_{0}$, where $\tilde{F}^{(0)}(\phi _{0})=h(y_{0})$. Near this
point, expanding relevant quantities, we get that in general $\phi -\phi
_{0}\backsim \sqrt{y_{0}-y}$, $R\backsim (y-y_{0})^{-3/2}$ diverges.

b) If $B=-\left| B\right| <0$, the function $h(y)\,$has a maximum $h_{m}$.
Solutions exist if $h_{m}>\tilde{F}(\phi _{0})$, in which case the
admissible region is bounded by two singularities: $y_{1}\leq y\leq y_{2}$,
where $\tilde{F}(\phi _{0})=h(y_{1,2})$.

Thus, in both cases a), b) the branch with the flat right infinity, typical
of the case $D=1$, now disappears..

3) $d<0$. The function $\tilde{F}^{(0)}(\phi )$ is monotonic. Then, the
branch with a linear dilaton vacuum $\phi =-y$ at right infinity, typical of
the case $D=1$, does not longer exists. Now at $y\rightarrow \infty $ the
dilaton $\phi =\left| d\right| ^{-1}\kappa ^{-1}e^{2y}$. The proper distance 
$l\backsim \int dy\sqrt{f}$ diverges in this limit, eq. (\ref{ry}) shows
that the curvature $R\backsim \exp (-\frac{2}{\left| d\right| \kappa }%
e^{2y}) $ is bounded. Thus, we have now a semi-infinite throat at $%
y\rightarrow \infty $ that has no counterpart in the case $D=1$.

\section{Degenerate models and extremal horizons}

The general form of solutions (\ref{fh}) implies that the parameters $\delta 
$, $\gamma $, $D$ entering (\ref{F}) are finite and nonzero. Meanwhile,
there exist some special (degenerate) cases, when the parameter $\delta $
vanishes. Then the form of the function $h$ in the right hand side of (\ref
{fh}) changes qualitatively. Below we discuss the corresponding cases
separately.. It is especially interesting that they contain, in particular,
extremal black holes ($T_{H}=0$). For comparison, let me recall that in
generic exactly solvable models considered in \cite{exact}, \cite{thr} such
solutions were absent. Extremal horizons with $\delta \neq 0$ were obtained
in \cite{ext} by expense of relaxing the finiteness of quantum stresses on
the horizon that needs imposing some severe restrictions on the form of the
coefficient $\tilde{F}$ near the horizon. Meanwhile, now (for degenerate
models) extremal horizons can be obtained, as will be seen, in a more
natural way.

\subsection{$\delta \rightarrow 0$, $\gamma \rightarrow 0$, $D$ is finite}

Let not only $\delta \rightarrow 0$, but also $\gamma \rightarrow 0$ with $D$
remaining finite, so the potential $U_{eff}=0=$ $\bar{\lambda}$. It can be
achieved at $Q=0$, $U=0$ or at $Q^{2}=W_{0}$, see (\ref{l}). Then eq. (\ref
{F}) has a non-trivial limiting transition, if $\delta \thicksim \gamma ^{2}$%
: $\tilde{F}=a\sigma +C+D$, $a\equiv -\frac{\kappa \gamma ^{2}}{2\delta }$ .
After an obvious rescaling we have 
\begin{equation}
\tilde{F}=\sigma \text{,}  \label{td1}
\end{equation}
where $f$ can be found from (\ref{sec}) and (\ref{root}).

It is also instructive to see, how the solutions of this type can be
obtained in the Schwarzschild gauge (\ref{gauge}) which was used in \cite
{ext} to realize, why the models under discussion do not fall into the
classes discussed in \cite{exact}. It follows from the integration of eq. (%
\ref{4}) in the Schwarzschild gauge (\ref{gauge}) that 
\begin{equation}
f=e^{-\psi }\chi \text{, }\chi =A\int_{\phi _{h}}^{\phi }d\phi ^{\prime }%
\tilde{F}^{\prime }(\phi ^{\prime })e^{-\psi }+A_{0}\text{.}  \label{2v}
\end{equation}
It was assumed in \cite{exact} that $A\neq 0$, $A_{0}=0$ that corresponds to
the horizon at $\phi =\phi _{h}$, when $\chi =f=0$, $\psi $ being finite on
the horizon. Meanwhile, another type of black hole solutions with a regular
horizon is also possible, if $A=0$, $A_{0}\neq 0$, $\psi \rightarrow \infty $
on the horizon. Then

\begin{equation}
f=f_{0}e^{-\psi }\text{, }\psi =\int d\phi \omega _{eff}  \label{sec}
\end{equation}
\begin{equation}
x=x_{0}\mu \text{, }\mu ^{\prime }=\tilde{F}^{\prime }e^{-\psi }\text{.}
\label{mx}
\end{equation}

As now $U_{eff}=0$, the definition $\omega _{eff}=U_{eff}^{\prime }/U_{eff}$
loses its sense but the formula (\ref{vex}) for the condition of exact
solvability is still valid. It changes its meaning: due to the condition $%
U_{eff}=0$ we may impose the dependences $V(\phi )$ and $u(\phi )$ at our
will, find $\omega $ from (\ref{vex}) and obtain $\psi ^{\prime }$ from (\ref
{sec}): 
\begin{equation}
\psi _{\pm }^{\prime }=\frac{u}{\kappa }\pm \sqrt{(\frac{u}{\kappa })^{2}-%
\frac{2V}{\kappa }}\text{.}  \label{root}
\end{equation}
If we want to have a well-defined classical limit $\kappa \rightarrow 0$, we
should take $\psi ^{\prime }=\psi _{-}^{\prime }=\omega _{eff}$.

Taking into account that the Schwarzschild and conformal gauge are related
according to $dx=d\sigma f$, one can establish easily the equivalence of
formulas (\ref{td1}) and (\ref{sec}) - (\ref{root}).

Consider explicit examples. Let $\omega _{eff}=const\equiv \omega _{1}>0$, $%
\tilde{F}=e^{\omega _{2}\phi }+\alpha \phi $, $V=\omega _{1}\omega
_{2}e^{\omega _{2}\phi }+\alpha \omega _{1}+\kappa \frac{\omega _{1}^{2}}{2}$%
, $\omega _{2}$ is a constant. Then $f=f_{0}e^{-\omega _{1}\phi }$, $%
x=x_{1}\exp [(\omega _{2}-\omega _{1})\phi ]+x_{2}e^{-\omega _{1}\phi
}+x_{h} $, where $x_{1}$, $x_{2}$, $f_{0}$ are another constants, the
quantity $\psi =\omega _{1}\phi $. Near the horizon $\phi \rightarrow \infty 
$ and $\psi $ diverges in agreement with the statement made above. If $%
\omega _{2}=\omega _{1}/2$, $f\backsim (x-x_{h})^{2}$ and we get the extreme
black hole. At infinity we have $f\backsim x$, $R\rightarrow 0$, thus this
extreme black hole is accelerated.

If $\omega _{2}=\omega _{1}\frac{(n-1)}{n}$, $f\backsim (x-x_{h})^{n}$.
Thus, for $n>2$ the horizon is ''ultraextreme'' in the sense that not only $%
f(x_{h})=0$, but also $f^{\prime }(x_{h})=0$.

If $\omega _{2}<0$, we have near the horizon $f\backsim (x-x_{h})$, so the
black hole is nonextreme. In particular, if $\omega _{2}=-\omega _{1}$, the
expansion has the form $f=a_{1}(x-x_{h})+a_{2}(x-x_{h})^{2}$, so the horizon
is regular, $R(x_{h})=-2a_{2}$.

Thus, depending on the properties of $\omega (\phi )$, $F(\phi )$ and $%
V(\phi )$, we have a diversity of possibilities, including extreme black
holes ($T_{H}=0$). This is in sharp contrast with exactly solvable models in 
\cite{exact}, where the Hawking temperature $T_{H}=const\neq 0$.

\subsection{$\delta \rightarrow 0$, $D\rightarrow -\infty $, $\gamma \neq 0$.
}

Let $C=C^{\prime }-D$, where $C^{\prime }$ is finite, 
\begin{equation}
D=D_{0}+D_{1}\delta ^{-1}+D_{2}\delta ^{-2}\text{,}
\end{equation}
where $D_{0}$, $D_{1}$ and $D_{2}$ are some constants. Eq. (\ref{F}) has a
well defined limit $\delta \rightarrow 0$, provided the coefficient $D_{2}$
is chosen to kill the divergent terms:$D_{2}=-\frac{\kappa \gamma ^{2}}{2}$.
Then, getting rid of the terms, linear in $\sigma $, by a simple shift, and,
for definiteness, taking the constant term positive, we obtain 
\begin{equation}
\tilde{F}^{(0)}=\kappa \frac{\gamma ^{2}}{4}(\sigma _{0}^{2}-\sigma ^{2})%
\text{.}  \label{d0}
\end{equation}
$\sigma _{0}^{2}=const$. It is seen that the existence of the degenerate
case is due to quantum effects only ($\kappa \neq 0$)$.$

From the trace equation 
\begin{equation}
U_{eff}=-\frac{\kappa \gamma ^{2}}{2f}\text{, }f\backsim W\text{.}
\label{uf}
\end{equation}

Consider an example. 
\begin{equation}
W=e^{2\phi }\text{, }F=e^{2\phi }-\alpha e^{-2\phi }+\beta \phi \text{, }%
\tilde{F}^{(0)}=e^{2\phi }-\alpha e^{-2\phi }+\tilde{\beta}\phi \text{, }%
\tilde{\beta}=\beta +2\kappa \text{, }\omega =-2\text{, }\alpha >0\text{.}
\label{exs2}
\end{equation}
Then 
\begin{equation}
f\backsim e^{2\phi }\text{.}
\end{equation}
There are two horizons at $\sigma =\pm \infty $, $\phi \rightarrow -\infty $%
, near which 
\begin{equation}
f\backsim \sigma ^{-2}\backsim (x-x_{h})^{2}\text{,}
\end{equation}
where $x$ is the Schwarzschild coordinate. Thus, the solution represent an
extreme black holes with two symmetric horizons. It is also clear that in
the case $\tilde{\beta}=0$ the fractional correction to $f$ have the order $%
\sigma ^{-2}$, so near the horizon $f=A_{1}(x-x_{h})^{2}+A_{2}(x-x_{h})^{4}$%
. It is known that the energy measured by a free falling observer is
proportional to $\frac{f^{\prime \prime \prime }}{f^{\prime }}$ \cite{triv}, 
\cite{and}. Thus, this quantity is finite and there are no thermal
divergencies on the event horizon, in contrast to a generic case.

In another case, if $\tilde{F}$ in (\ref{exs2}) contains the term $%
ce^{-3\phi }$, with $c<0$, near the horizon we have $f$ $\backsim \sigma
^{-4/3}\backsim (x-x_{h})^{4}$. Thus, a black hole is ''ultraextreme''.

\section{Generic extremal semiclassical black holes}

In the preceding sections we mainly considered exactly solvable models and
found that only in some special limits one can hope to find for such models
solutions of the extremal black hole type. Thus, extremality and exact
solvability rather seldom come into play simultaneously. Meanwhile, the
issue of extremal black holes is important by itself and, therefore, it is
of interest to elucidate, what information about corresponding solutions can
be extracted from basic equation without severe restriction to exactly
solvable models only.

In what follows we are interested in such extremal black holes that quantum
stresses remain bounded on a horizon. In principle, regular extremal
horizons may exist even if this restriction is relaxed \cite{ext}, \cite{mod}
but we do not discuss here these rather special cases and assume that, as
usual, a static extreme black hole with a regular horizon realizes the
Hartle-Hawking state with the temperature $T=T_{H}=0$.

It is seen from eqs. (\ref{8}) and (\ref{ps}) that the expression for $%
T_{1}^{1(PL)}$ can be rewritten, as 
\begin{equation}
T_{1}^{1(PL)}=-\frac{\kappa }{4\pi f}[A^{2}-f_{x}^{\prime 2}]\text{. }
\label{t11}
\end{equation}

As for extremal horizons $f_{x}^{\prime }(x_{h})=0$, the condition of
regularity demands that $A=0$. Then, according to (\ref{ps}), 
\begin{equation}
f=f_{0}e^{-\psi }\text{,}  \label{fps}
\end{equation}
where $f_{0}$ is a constant. (Recall that in \cite{exact} we assumed that $%
\psi $ was regular on the horizon and took the quantity $A=f^{\prime
}(x_{h})\neq 0$ in agreement with the non-extreme character of black holes
considered in \cite{exact}.)

One can rewrite (\ref{sc11}) and (\ref{dif}) as 
\begin{equation}
\left( V-\psi ^{\prime }u+\kappa \frac{\psi ^{\prime 2}}{2}\right)
z^{2}=U_{eff}e^{\psi }\text{,}  \label{e1}
\end{equation}
\begin{equation}
C_{1}z+C_{2}\frac{\partial z}{\partial \phi }=0\text{,}  \label{e2}
\end{equation}
\begin{equation}
C_{1}=u^{\prime }-V+\kappa (\frac{\psi ^{\prime 2}}{2}-\psi ^{\prime \prime
})\text{, }C_{2}=(u-\kappa \psi ^{\prime })  \label{e3}
\end{equation}
(recall that $z\equiv \frac{\partial \phi }{\partial x}$).

For exactly solvable models, when eq.(\ref{v0}) is satisfied, the left hand
side of eq. (\ref{e1}) vanishes and we return to the degenerate case $%
U_{eff}=0$, considered above - see eqs. (\ref{2v}) - (\ref{root}). From now
on, we assume that $U_{eff}\neq 0.$ This means that the condition (\ref{v0})
is abandoned.

Let $z\equiv \sqrt{Y(\phi )}$, $Y\equiv \frac{a}{b}$. We want to derive from
(\ref{e1}), (\ref{e2}) the closed equation for the single quantity $\psi $.
We have 
\begin{equation}
2C_{1}+C_{2}(\frac{a^{\prime }}{a}-\frac{b^{\prime }}{b})=0\text{.}
\end{equation}
Here $a=U_{eff}e^{\psi }$, $b=V-\psi ^{\prime }u+\kappa \frac{\psi ^{\prime
2}}{2}$ (henceforth we write down for shortness $U$ instead of $U_{eff}$).
After some manipulations we get the equation ($y\equiv \psi ^{\prime }$) 
\begin{equation}
y^{\prime }d\text{ }+d_{3}y^{3}+d_{2}y^{2}+d_{1}y+d_{0}=0\text{,}  \label{yd}
\end{equation}
\begin{equation}
d=(u^{2}-2V\kappa )\text{,}
\end{equation}

\begin{equation}
d_{3}=\frac{\kappa }{2}(u-\kappa \omega )
\end{equation}

\begin{equation}
d_{2}=-u^{2}+(\frac{3\omega u}{2}-V)\kappa
\end{equation}

\begin{equation}
d_{1}=u(3V-\omega u-u^{\prime })+\kappa (V^{\prime }-\omega V)
\end{equation}

\begin{equation}
d_{0}=V\left( 2u^{\prime }-2V+\omega u\right) -uV^{\prime }
\end{equation}

The classical case corresponds to $\kappa =0$. Then

$d=u^{2}$, $d_{3}=0$, $d_{2}=-u^{2}$, $d_{1}=u(3V-\omega u-u^{\prime }).$
Let $V\equiv u\alpha $. Then eq. (\ref{yd}) turns into 
\begin{equation}
y^{\prime }-y^{2}+y(3\alpha -\omega -\frac{u^{\prime }}{u})+\alpha (\frac{%
u^{\prime }}{u}-2\alpha +\omega )-\alpha ^{\prime }=0\text{.}  \label{ycl}
\end{equation}

By substitution $y=\eta +\alpha $ we get 
\begin{equation}
\eta ^{\prime }-\eta ^{2}+X\eta =0\text{, }X\equiv (\alpha -\omega -\frac{%
u^{\prime }}{u})\text{.}
\end{equation}
\begin{equation}
\eta \equiv -\frac{\chi ^{\prime }}{\chi }\text{.}
\end{equation}
\begin{equation}
\chi ^{\prime \prime }+\chi ^{\prime }X=0\text{.}
\end{equation}
\begin{equation}
\chi ^{\prime }=\exp [-\int d\phi Y(\phi )]\text{.}
\end{equation}
This coincides with eqs. (\ref{ff}). - (\ref{coor}), in which $\kappa =0$ is
put in the right hand sides.

Consider now the quantum case. Following the procedure of \cite{kunst1}, we
can exclude the kinetic term and take the coefficient $F$ as a new dilaton
field, provided $F(\phi )$ is monotonic: $V=0$, $F(\phi )\Longrightarrow
\phi $. Then 
\begin{equation}
d=1\text{, }d_{3}=\frac{\kappa (1-\kappa \omega )}{2}\text{, }d_{2}=-1+\frac{%
3\omega \kappa }{2}\text{, }d_{1}=-\omega \text{, }d_{0}=0\text{.}
\end{equation}
We get 
\begin{equation}
y^{\prime }+\frac{\kappa (1-\kappa \omega )}{2}y^{3}+\left( -1+\frac{3\omega
\kappa }{2}\right) y^{2}-\omega y=0\text{.}  \label{adj}
\end{equation}

If we want to have $y=-\frac{2}{\phi }$ (zero temperature Bertotti-Robinson
spacetime), we obtain \cite{deg}, \cite{fab} 
\begin{equation}
\omega =\frac{1}{\phi +\kappa }\text{, }U=U_{0}(\phi +\kappa )\text{.}
\label{u2}
\end{equation}

In principle, we may find $\omega $ in terms of $y$ and $y^{\prime }$ from (%
\ref{adj}) and adjust $\omega $ and $U$ to have any desired behavior $y(\phi
)$, typical of an extremal black hole. This means that equality $y=-2/\phi $
should be fulfilled near the horizon only, so the curvature does not need to
be constant everywhere. In turn, it means that, on the basis of eq. (\ref{yd}%
), one can describe diversity of models with desired features. To some
extent, the procedure under discussion (adjusting the potential to fixed
solutions or properties) is similar to finding the stress-energy tensor $%
T_{\mu \nu }$ from geometry of spacetime, using Einstein equations. In
general relativity, such a procedure is not very natural since it leaves the
question about physical nature of $T_{\mu \nu }$ unresolved. However, for 2D
models it looks more reasonable since we simply fixed the form of the
potential. Apart from this, in this approach one is able to describe
geometries of physical interest, including, for-example, the potential (\ref
{u2}), relevant for self-consistent scenario of black-hole evaporation \cite
{fab}.

\subsection{Extremal horizons with and without divergencies of quantum
stresses}

As is known, in the extreme case the coordinate dependence of a dilaton near
the horizon in the Schwarzschild gauge is non-analytical, that is connected
with the appearance of divergencies of the quantum stresses on the horizon
in the reference frame of a free falling observer. This was shown in \cite
{triv} for a particular 2D model, obtained by spherically-symmetrical
reduction from 4D Lagrangian. It is now instructive to trace, on the basis
of eq. (\ref{yd}), how these divergencies appear in the general case. Let a
horizon be situated at $\phi =0$, where $y\rightarrow \infty $ according to (%
\ref{fps}). Dividing eq. (\ref{yd}) by $y^{3}$ and taking the limit $%
y\rightarrow \infty $, we find that if we want this limit indeed to
correspond to the horizon, $d_{3}\rightarrow 0$, $\omega \rightarrow
u/\kappa $ on the horizon. Expanding $d_{3}$ in terms of $\phi $ and
substituting the asymptotic expression 
\begin{equation}
y=-\frac{\varepsilon }{\phi }\text{, }\varepsilon \equiv \frac{2}{1+\tau }%
\text{,}  \label{et}
\end{equation}
into (\ref{yd}), we get, equating terms at the $\phi ^{-2}$: 
\begin{equation}
\tau ^{2}+3\tau -2\kappa \nu =0\text{,}  \label{tau}
\end{equation}
\begin{equation}
\nu \equiv \lim_{\phi \rightarrow 0}\left( \frac{2V+u^{\prime }-\frac{%
U^{\prime \prime }}{U^{\prime }}\kappa }{u^{2}-2V\kappa }\right) \text{.}
\label{n}
\end{equation}

Eqs. (\ref{fps}), (\ref{et}) give us the asymptotic form $f=\phi
^{\varepsilon }$ near the horizon. Comparing it with the coordinate
asymptotics $f=const(x-x_{h})^{2}$, typical of an extremal horizon, we get $%
\phi \backsim (x-x_{h})^{1+\tau }$. In the classical limit $\kappa
\rightarrow 0$ $\varepsilon =2$ and dilaton dependence of the metric $%
f=const\phi ^{2}$ becomes analytical: $f=const\phi ^{2}$.

In the model considered in \cite{triv} $F=e^{-2\phi }$, $u=-2e^{-2\phi }$, $%
U=2-2Q^{2}e^{2\phi }$, $V=2e^{-2\phi }$, $e^{-2\phi _{h}}\equiv d_{h}$.$\,$%
Then (\ref{n}) gives us 
\begin{equation}
\nu =\frac{3}{d_{h}-\kappa }
\end{equation}
in full accord with \cite{triv}.$^{{}}$

If $\nu =0$, we get $\tau =0$ - divergencies under discussion do not appear
on the horizon (the root $\tau =-3$ is not suitable since we must have $%
\varepsilon >0$ to get the horizon at $\phi =0$). For example, this takes
place for the model $F=e^{-\phi }$, $V=2e^{-\phi }$, $U=U_{1}+U_{2}e^{-5\phi
}$.

\subsection{Ultraextreme black holes}

It was noticed in Sec. V that 2D dilaton gravity admits such a kind of black
holes that near the horizon 
\begin{equation}
f\backsim (x-x_{h})^{n}\text{, }n>2  \label{bhn}
\end{equation}
(''ultraextreme'' black holes). Now we will show that account for quantum
backreaction changes this result drastically.

It follows from (\ref{e2}) that 
\begin{equation}
\frac{\partial \ln z}{\partial \phi }=\frac{\kappa (\frac{y^{2}}{2}%
-y^{\prime })+u^{\prime }-V}{\kappa y-V}\text{,}  \label{lnz}
\end{equation}
whence in the limit $y\rightarrow \infty $, corresponding to the horizon, we
obtain 
\begin{equation}
\frac{\partial \ln z}{\partial \phi }=\frac{y}{2}-\frac{y^{\prime }}{y}+...%
\text{.}  \label{y2}
\end{equation}
On the other hand, comparing the asymptotic behavior $f\backsim
(x-x_{h})^{n} $ to the general formula (\ref{fps}), we find that near the
horizon 
\begin{equation}
\frac{\partial \ln z}{\partial \phi }=\frac{y}{n}-\frac{y^{\prime }}{y}+...%
\text{.}  \label{yn}
\end{equation}

Comparing (\ref{y2}) and (\ref{yn}), we see that $n=2$. Thus, ultraextreme
black holes are impossible if quantum effects are taken into account.

It is instructive to trace the origin of discrepancy between the classical
and quantum cases. If the quantum-coupling parameter $\kappa =0$ from the
very beginning, it is seen from (\ref{lnz}) that $\frac{\partial \ln z}{%
\partial \phi }$ remains bounded on the horizon, so do $z$ and $z^{\prime }$%
. Then one can infer from (\ref{e1}) (where the terms with $\kappa $ are
dropped) and (\ref{fps}) that for a black hole of the type (\ref{bhn}) the
potential behaves near the horizon like $U\backsim f^{\prime }\backsim
(x-x_{h})^{n-1}\backsim (\phi -\phi _{h})^{n-1}$. In turn, this means that $%
\omega =(n-1)/\phi $. Then one can check that the solution under discussion
obeys eq. (\ref{ycl}), where $y=-n/\phi $ near the horizon.

On the other hand, it is just terms with $\kappa $ (which are absent for a
classical system) which dominate both the numerator and denominator in (\ref
{lnz}) and in eq. (\ref{e1}) since they contain $y$ and $y^{\prime }$ and
this is just the reason why the structure of solutions changes qualitatively.

It is worth noting that in general relativity the issue of influence of
quantum backreaction on the character of solutions is highly non-trivial,
because of absence of exact solutions and mathematical complexities. It was
shown in \cite{lowe}, \cite{jo} that in general relativity the presence of
quantum backreaction is compatible with the existence of extremal black
holes. The similar conclusion was reached in \cite{barb} for 2D theories
with non-conformal scalar fields. However, as we saw above, the situation
looks differently for ultraextreme solutions: quantum effects forbid them.

The statement about impossibility of ultraextreme semiclassical black hole
refer to the generic case $U\neq 0$. However, if $U=0$, such solutions do
become possible as we saw in some examples in Sec. VI following eq. (\ref
{root}).

\section{Summary}

The condition of exact solvability in 2D dilaton semiclassical gravity (\ref
{v0}) involves, via the effective potential, the value of an electric
charge. This makes it, in general, impossible to generalize these conditions
to have exact solutions for {\it any} value of charge. This is in sharp
contrast, say, with general relativity where the Reissner-Nordst\"{o}m
solution is known for an arbitrary charge. It turned out that there are only
two ways to circumvent the obstacle under discussion and exclude the charge
from $U_{eff}$ that leads to two possibilities. In the first case the
solutions look very much like uncharged ones but with new parameters which
contain the electric charge. In the second case the original potential $U=0$%
, and $U_{eff}$ is due to the dilaton-Maxwell coupling entirely. Again, the
structure of exact solutions follows from general approach developed for the
uncharged case \cite{exact} but now the amplitude of the effective potential
is negative. First, it affects the structure of spacetime, as compared to
the case $U_{eff}>0$, typical of uncharged solutions. Second, it admits the
special ''degenerate'' class of exactly solvable models which is absent in
the case of potentials with a positive amplitude.

The relationship between the generic exactly solvable case and its special
degenerate subclass is of interest also on its own, independent of the
presence of an electric charge. We described this relationships in terms of
some universal function $q(\sigma )$, where $\sigma $ is a conformal
coordinate. Some non-trivial limiting transitions change its character
completely and show that there exist exact solutions qualitatively different
from those described in \cite{exact}. The main new feature here is that
these exact solutions may include extreme black holes.

On the other hand, usual (non-degenerate) exactly solvable models admit only
non-extreme black holes. Therefore, the attempt to find extreme black holes
beyond the degenerate subfamilies of exactly solvable models forced us to
relax the condition (\ref{v0}) of exact solvability at all and consider
generic extreme black holes ($U_{eff}\neq 0$)\ in the Hartle-Hawking state.

It is instructive to summarize in a table, in which cases 2D semiclassical
dilaton theories admit extreme black holes.

\begin{tabular}{|l|l|l|l|}
\hline
& $\delta =0$; $\gamma $, $A$, $U\neq 0$ & $\delta ,\gamma ,A,U=0$ & $A=0$, $%
U\neq 0$ \\ \hline
Exactly solvable & degenerate subclass & degenerate subclass & $-$ \\ \hline
$q(\sigma )$ & $\frac{\kappa \gamma ^{2}}{4}(\sigma _{0}^{2}-\sigma ^{2})$ & 
$\sigma $ & $-$ \\ \hline
EBH & admissible & admissible & mandatory \\ \hline
Divergencies & $+$ & $-$ & $-$ \\ \hline
UEBH & admissible & admissible & forbidden \\ \hline
\end{tabular}

Here EBH and UEBH mean extreme and ultraextreme black holes. In the last but
one row we indicated, whether or not quantum stresses on a horizon diverge
(that is determined by the coefficient $A$ in (\ref{t11}). To avoid possible
confusion, we want to pay attention that information in the table concerns
divergencies in the original (static) reference frame, whereas those found
in \cite{triv} refer to a free falling observer. The fact, that divergencies
of the first kind are compatible with finiteness of curvature, gives us one
more example (apart from those in \cite{ext}, \cite{mod}) that in dilaton
gravity infinite backreaction may be compatible with a regular horizon.

We have managed to derive one closed equation with respect to the auxiliary
field $\psi (\phi )$ which governs the behavior of the system. In
particular, with the help of this equation we analyzed in a general setting
divergencies of quantum stresses on the horizon \cite{triv}, \cite{and} and
found the conditions when these divergencies completely cancel.

In a sense, it is rather hard to combine exact solvability with extremal
horizons: only degenerate cases of the general scheme \cite{exact} admit
such a combination. However, information which can be extracted from the
equation for $\psi (\phi )$, enables us to find exact solutions in another,
more restricted sense: choosing any desirable behavior of the metric as a
function of dilaton, we restore the Lagrangians which ensure such a behavior
(''inverse problem'').

We also found and analyzed exact solutions without the condition of exact
solvability for a special branch of solutions with the constant dilaton
field. It is shown that the correspondence between these solutions and
extremal horizon of the main branch remains intact in spite of the presence
of backreaction, but with the reservation that quantum stresses remain
bounded on the horizon. If the later condition is relaxed, there exist
regular extremal horizons having no counterparts among solutions with a
constant dilaton field.

As far as the role of quantum backreaction is concerned, it turns out also
that, although it does not affect the very existence of extremal black
holes, it forbids their ultraextreme versions (except some special cases).
This stimulates to examine this issue further for a more realistic 4D case,
where such solutions are known on the classical level - for instance, among
Reissner-Nordstr\"{o}m-de Sitter solutions.





%
%

%
%


\begin{references}
\bibitem{callan}  C.G. Callan, S. Giddings, J.A. Harvey J. A and A.
Strominger, Phys. Rev. D\ 45 (1992) R1005;

T. Banks, A. Dabholkar, M. R. Douglas, and M. O. 'Loughlin, Phys. Rev.\ D\
45 (1992 ) 3607.

\bibitem{od}  S.Nojiri and S. Odintsov, Int. J. Mod. Phys. A 16 (2001 ) 1015.

\bibitem{dv}  D. Grumiller, W. Kummer and D. V. Vasilevich, Phys.Rept. 369
(2002) 327.

\bibitem{fil1}  A. T. Fillipov and V. G. Ivanov, Phys.Atom.Nucl. 61 (1998)
1639 (hep-th/9803059).

\bibitem{fil2}  A. T. Fillipov, Mod.Phys.Lett. A11 (1996) 1691.

\bibitem{eliz2}  E. Elizalde and S.D. Odintsov, Nucl. Phys. B399 (1993) 581.

\bibitem{eliz4}  E. Elizalde, P. Fosalba-Vela, S. Naftulin, S. D. Odintsov,
Phys. Lett. B 352 (1995) 235.

\bibitem{strobl}  T. Kloesch and T. Strobl, Class. Class.Quant.Grav. 13
(1996) 965-984; Erratum-ibid. 14 (1997) 825.

\bibitem{pelzer}  H Pelzer and T. Strobl, Class.Quant.Grav. 15 (1998) 3803.

\bibitem{rst}  J. G. Russo, L. Susskind, and L. Thorlacius, Phys. Rev. D{\bf %
\ }46 (1992) 3444; Phys. Rev. D{\bf \ }47 (1992) 533.

\bibitem{bil}  A. Bilal and C. G. Callan, Nucl. Phys. B 394, 73 (1993).

\bibitem{alw}  S. P. de Alwis, Phys. Rev. D 46 (1992) 5429.

\bibitem{rob}  G. Michaud and R. C. Myers, Two-Dimensional Dilaton Black
Holes, gr-qc/9508063.

\bibitem{bose}  S. Bose, L. Parker and Y. Peleg, Phys. Rev. D 52 (1995) 3512.

\bibitem{kaz}  Y. Kazama, Y. Satoh, and A. Tsuichiya, Phys. Rev. D 51 (1995)
4265.

\bibitem{exact}  O. B. Zaslavskii, Phys. Rev. D 59 (1999) 084013.

\bibitem{polyakov}  A. M. Polyakov, Phys. Lett. B 103 (1981) 207.

\bibitem{solod}  S. N. Solodukhin, Phys. Rev. D{\bf \ }53 (1996) 824.

\bibitem{israel}  V. P. Frolov, W. Israel, and S. N. Solodukhin, Phys. Rev. D%
{\bf \ }54 (1996) 2732.

\bibitem{frolov}  V. P. Frolov, Phys. Rev. D{\bf \ }46 (1992) 5383.

\bibitem{const}  O. B. Zaslavskii, Phys. Lett.{\bf \ }B{\bf \ }424 (1998)
271.

\bibitem{deg}  J. Cruz, A. Fabbri, D. J. Navarro and J. Navarro-Salas, Phys.
Rev. D 61 (1999) 024011.

\bibitem{and}  D.J. Loranz, W.A. Hiscock and P.R. Anderson, Phys. Rev. D 52
(1995) 4554.

\bibitem{gibper}  G.W.Gibbons and M.J.Perry, Int.J.Mod.Phys.{\bf \ }D{\bf \ }%
1{\bf \ }(1992) 335.

\bibitem{nappi}  C. R. Nappi and A. Pasquinucci, Mod. Phys. Lett. A 7 (1992)
3337.

\bibitem{gukov}  N. Berkovits, S. Gukov and B.C. Vallilo Nucl.Phys. B 614
(2001) 195.

\bibitem{kunst1}  J. Gegenberg, G. Kunstatter, D. Louis-Martinez, Phys. Rev.
D 51 (1995) 1781.

\bibitem{cad}  S. Monni and M. Cadoni, Class. Quant. Grav. 14, (1997) 517.

\bibitem{park}  Y. Kiem, C.-Y. Lee and D. Park, Class. Quant. Grav. 15
(1998) 2973.

\bibitem{thr}  O.B. Zaslavskii, Phys. Lett. B 459 (1999) 105.

\bibitem{nonext}  O.B. Zaslavskii, Phys. Rev. D 61 (2000) 064002.

\bibitem{cruz}  J. Cruz and J. Navarro-Salas, Phys. Lett. B 375 (1996) 47.

\bibitem{ext}  O.B. Zaslavskii, Phys. Lett. B 475 (1999 ) 33.

\bibitem{mod}  O. B. Zaslavskii, Entropy of semiclassical 2D dilaton black
holes away from the Hawking temperature, hep-th/0208206 (To be published in
Mod. Phys. Lett. A).

\bibitem{triv}  S. Trivedi, Phys. Rev. D 47 (1993) 4233.

\bibitem{fab}  A. Fabbri, D. J. Navarro and J. Navarro-Salas, Phys. Rev.
Lett. 85 (2000) 2434.

\bibitem{lowe}  D. A. Lowe, Phys. Rev. Lett. 87 (2001) 029001.

\bibitem{jo}  J. Matyjasek and O. B. Zaslavskii, Phys. Rev. D 64 (2001)
104018.

\bibitem{barb}  C. Barbachoux and A. Fabbri, Phys. Rev. D 66 (2002) 024012.
\end{references}
\end{document}